\documentclass[aps,prl,showpacs,unsortedaddress,twocolumn]{revtex4-1}        
\usepackage{amsmath}
\usepackage{subfig}
\usepackage{dcolumn}
\usepackage{multirow}
\usepackage{graphicx}
\usepackage{bbold}
\usepackage{epsf}
\usepackage{epstopdf}
\usepackage{color}

\begin{document}
\title{Tunability of the Berry phase in gapped graphene}
\author{Andrea Urru} 
\affiliation{Dipartimento di Fisica dell'Universit\`a di Cagliari, Cittadella Universitaria, Monserrato, I-09042 Cagliari, Italy}\author{Giulio Cocco}
\affiliation{Universit\"at Freiburg, Freiburg im Breisgau, Germany}
\author{Vincenzo Fiorentini}
\affiliation{Dipartimento di Fisica dell'Universit\`a di Cagliari and CNR-IOM, Cittadella Universitaria, Monserrato, I-09042 Cagliari, Italy}
\date{\today}
\begin{abstract}
When a gap of tunable size opens at the conic band intersections of graphene, the Berry phase  does not vanish abruptly, but progressively decreases as the gap increases. The phase depends on the reciprocal-space path radius, i.e., for a doped system, the Fermi wave vector. The phase and its observable consequences can thus be tuned continuously via gap opening  --by a modulating potential induced by strain, epitaxy, or nanostructuration-- and doping adjustment.
  \end{abstract}
\maketitle

A characteristic feature of graphene is its  cone-shaped singular band     dispersion near the six vertexes of the Brillouin zone  at  the Fermi energy \cite{wallace,castroneto}. The locally linear  dispersion relation enables a description  of near-Fermi-level carriers  in graphene as massless Dirac quasi-particles \cite{castroneto,dassarma}. Limiting for simplicity the discussion to electrons, and to the vicinity of one of the two types of  vertexes (for example, the point $K$), the  periodic part of the Bloch wavefunction for wavevector {\bf q} is
\begin{equation}
u^{K}_{\bf q} (\mathbf{r}) = \frac{1}{\sqrt{2}} 
\begin{pmatrix}
{1} \\ {\rm e}^{i \Theta_{\bf q}}
\end{pmatrix}
\label{diracwf}
\end{equation}
with $\Theta_{\bf q}$=$\arctan{({q_x}/{q_y})}$.
We now recall that the Berry phase  \cite{berry} 
\begin{equation}
\Gamma=\oint_{\Lambda} {\bf A}\cdot d{\bf \Lambda}=\int_{S}  {\bf B}\cdot d{\bf S}\label{phase}
\end{equation}
 is   the circulation on a line $\Lambda$ in $q$-space of the Berry connection
{\bf A}=$i$$\langle u^{K}_{\bf q}|\nabla_{\bf q}|u^{K}_{\bf q}\rangle$ or the flux of the Berry curvature {\bf B}=$\nabla$$\times${{\bf A}} on  the area $S$ encircled by $C$. It follows that an electron in graphene collects a  Berry phase of $\pi$ as it circles the singular (or Dirac, or conical) point $K$. By contrast, an electron in a 2-dimensional free-electron gas (2DEG),  with parabolic dispersion and plane-wave eigenfunction,  has a vanishing  Berry phase (modulo 2$\pi$ \cite{marzari}) as it circles its locally parabolic band extremum. 

The non-trivial phase in graphene is directly observable in the Shubnikov-de Haas oscillations in magnetic field and in the quantum Hall effect \cite{qheberry,qheberry2,expmanif}, which is, due to the very existence of this phase, quite different that in a 2DEG \cite{dassarma}.
The Dirac quasiparticles also enjoy an additional constant of motion (chirality or helicity) \cite{chirality} which is responsible for the suppression of backscattering known as the Klein paradox, itself a key ingredient of the exceptional transport properties of graphene. Chirality is a good quantum number as long as the Dirac hamiltonian is applicable, i.e. the  dispersion is  linear  \cite{castroneto}, and its  conservation is directly linked to the non-trivial phase; indeed, both are a consequence of the linear dispersion, and chirality eigenvalues and phase values are in a one-to-one correspondence.

A natural question is now what happens to the Berry phase when graphene is subjected to a weak periodic bandwidth-limited potential that opens a small gap at the Dirac points. By the demands of time reversal and analiticity \cite{bassani}, the band dispersion must now be  quadratic at the band gap edges, i.e. near the (former) Dirac point; however,  for a weak potential, the  linear dispersion is likely to survive over much of the region around the former conical intersection. 

In this paper we examine the Berry phase in the regime just outlined, when a gap of tunable size opens at the conic intersections. We find that the Berry phase does not vanish abruptly, but progressively decreases as the gap increases. The phase now depends on, and increases with, the radius of the path in $q$-space (when electrons are doped into the system, that radius is just the Fermi wave vector, which obviously can be directly tuned). It follows that the Berry phase and its observable consequences can be tuned away from their value in graphene by i) tuning the gap by a modulating potential, strain, or nanostructuration and ii) adjusting the doping level. 

From the practical viewpoint, our suggestion is easy to countercheck and exploit. Gaps from 20 meV upwards can be produced by strain \cite{gapopening3}, epitaxy \cite{gapopening2}, nanostructuration \cite{gapopening1}. Doping would be in the typical currently-reachable range. The main effect one expects would be a reduction of the extra phase in the quantum oscillations in a magnetic field. A worsening of transport properties via  the reduction of chirality is  also expected, although this may be harder to interpret being superimposed on the effects of gap opening, dispersion change, etc.

We study bands and  Berry curvature and phase for a minimal first-neighbor tight-binding model of graphene, using the PythTB suite by Coh and Vanderbilt \cite{pyth}. The lattice vectors are 
$a$(1,0) and $a$(1/2,$\sqrt{3}$/2), with site coordinates  (1/3,1/3) and (2/3,2/3).  The energies are henceforth in eV; the band structure of graphene is on the correct energy scale  setting the hopping parameter  to $t$=--3 in internal units.

As expected, we find that lowering the symmetry   of the lattice or of the 
first-neighbor hoppings does not open a gap, but only displaces the conical intersections within  $q$-space. 
A gap does open when different on-site tight-binding terms, e.g. $\pm$$\delta$, are added to the two sublattices, which are thus rendered inequivalent. This is the same as  assuming that graphene is subjected to an external modulating potential of amplitude $\Delta$=2$\delta$ and zero average with the periodicity of a sublattice, a concept investigated in terms of strain-induced pseudomagnetic fields by Guinea {\it et al.} \cite{gapopening0}. (See below for a discussion of  long-wavelength modulations.) The resulting gap  $E_g$$\simeq$$\Delta$ at the former conical intersections can therefore be considered an adjustable parameter. 

\begin{figure}[ht]
\centering
\includegraphics[width=.45\textwidth]{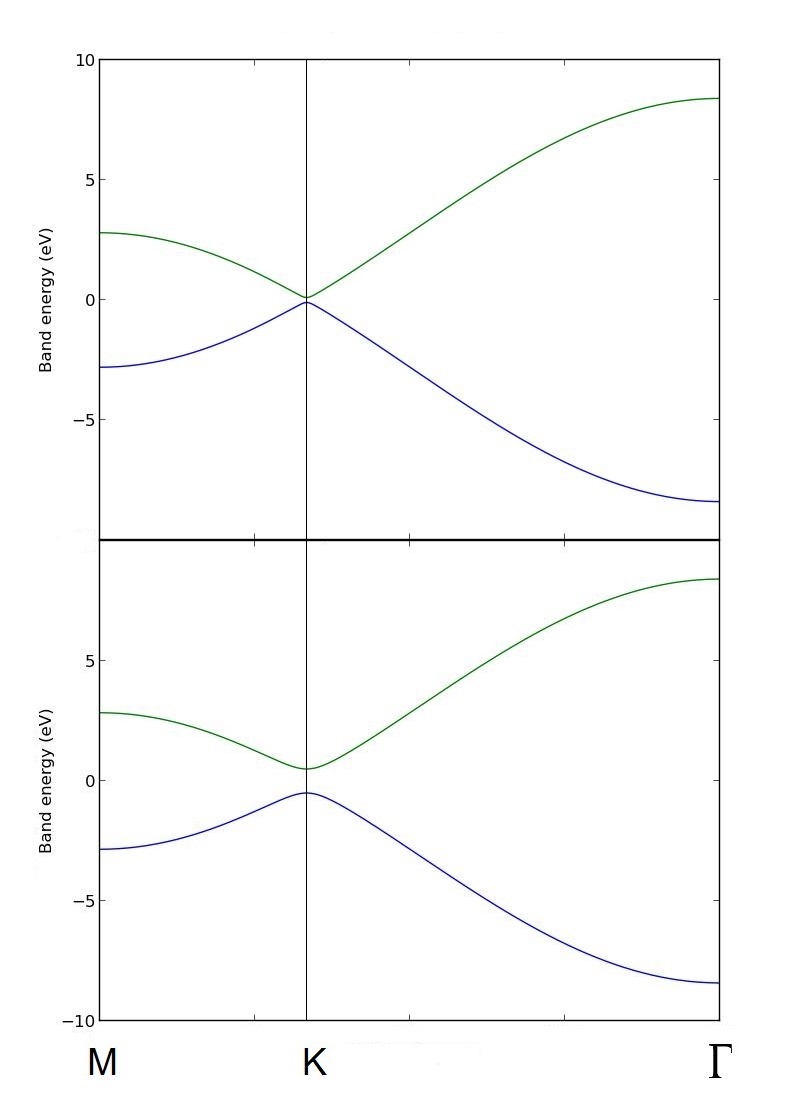}
\caption{Band stucture of graphene with on-site energies $\delta$=0.1 (top) and 0.5 eV (bottom).}
\label{fig2}
\end{figure}

As  the symmetry of two triangular sublattices is broken by turning on the amplitude $\Delta$, a gap appears at the Dirac point. This is shown in Fig. \ref{fig2} for two values of $\Delta$.  By analiticity and Kramers degeneracy, the top valence and bottom conduction bands, resulting from the anticrossing of the formerly-crossing conical bands, must have quadratic dispersion in the immediate vicinity of the former Dirac point. This has consequences, as we now discuss, for the Berry phase.

 In pristine graphene the phase is $\pi$ (as can be calculated numerically, modulo a minor technicality \cite{tech}). This property can be interpreted in various ways. Eq.\ref{phase}
and attendant definitions lend themselves naturally to an analogy  with  electrodynamics,  whereby the  connection is identified with the vector potential, the curvature with the magnetic field, and the  phase with the magnetic flux. Similarly to the electrodynamics case, the Berry curvature is  a gauge-invariant observable geometric property of the wavefunction in $q$-space; the Berry connection is physically measurable only when integrated, but it is nevertheless relevant in the present case, much in the same way as the vector potential in the Aharonov-Bohm effect.

In terms of this analogy, the non-trivial  Berry phase of graphene can be thought of as due to a singular curvature, i.e. the  phase is  the flux of the curvature at the   conical intersection point. Alternatively, one can see this as the electron  picking up the phase as the circulation of the Berry connection on  any path in $q$-space around the conical point: this is because the connection has the same form everywhere around the Dirac point if the dispersion is linear everywhere. Again, this is  akin to the Aharonov-Bohm effect: where the electrons circulate, the field (the curvature) is zero; not so the vector potential (the connection).  A possible magnetic analog of the Berry curvature is the  field in an extremely long solenoid of extremely small  radius, i.e. entirely confined within the solenoid and strictly zero outside (asymptotically for infinite length and infinitesimal radius), which still produces a finite flux over a sectional area. A color-coded contour plot of the Berry curvature would display zero, except for a $\delta$-like spike at the Dirac point \cite{tech}.

\begin{figure}
\centering
\includegraphics[width=.45\textwidth]{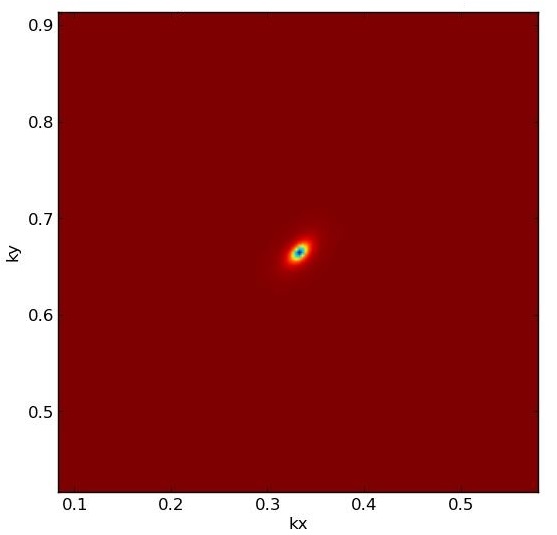}
\includegraphics[width=.45\textwidth]{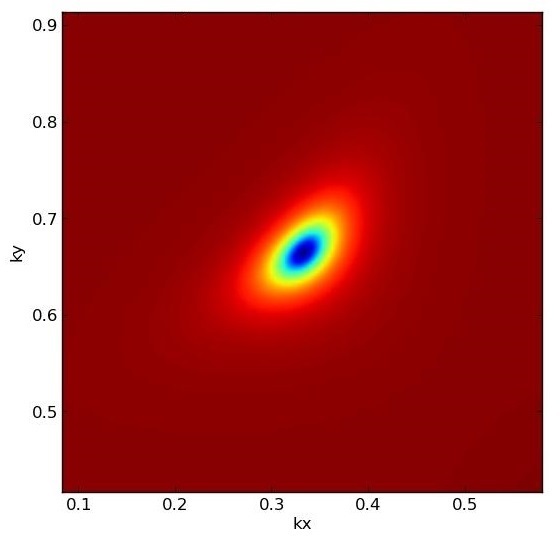}
\caption{(Color online) Berry curvature in $q$-space  obtained with the same on-site terms as in Fig.1.}
\label{fig3}
\end{figure}

As shown above, when we turn on the amplitude $\Delta$ a gap appears at the Dirac point, and a region with  quadratic dispersion appears in its  immediate vicinity. The Berry curvature, displayed in Fig.\ref{fig3} for two $\Delta$ values, is non-zero in a finite (and increasing with $\Delta$) region of $q$ space, and now looks more like the field generated by a finite-length, finite-radius solenoid. 
The  Berry curvature being non zero, we expect a non-zero Berry phase, which will depend on the radius of the path that the electrons circle upon. Indeed we find that as the radius increases, the phase value rapidly saturates. The resulting  value is {\it always smaller than $\pi$ and greater than 0 whenever $\Delta$ is non-zero}.  
Upon doping or field-effect injection, the path radius just mentioned is in effect the Fermi wavevector $q_F$, which can therefore be tuned via the free-carrier density $n_{\rm 2D}$=$q_F^2$/2$\pi$ in the conduction band. Therefore,  the Berry phase can be tuned in a substantial range using two knobs: $\Delta$ (the external potential)  and the path radius (or $q_F$, or $n_{\rm 2D}$; doping or injection).

\begin{figure}[h]
\centering
\includegraphics[width=.5\textwidth]{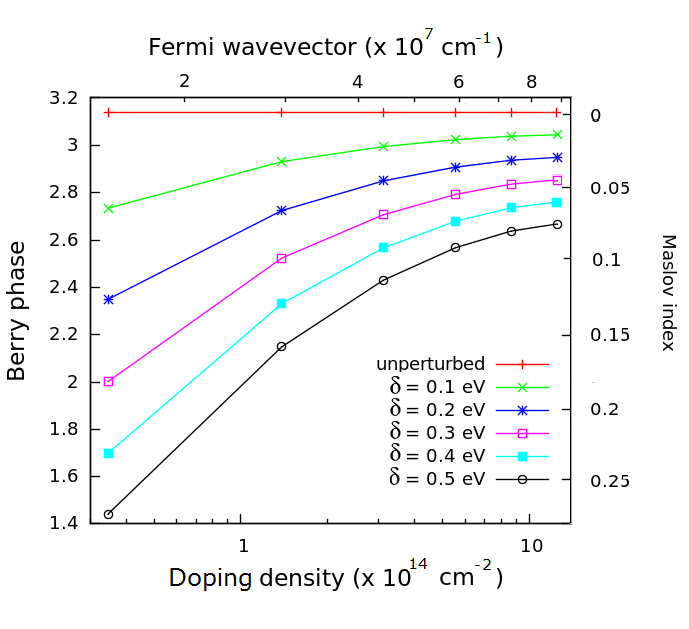}\\\includegraphics[width=.5\textwidth]{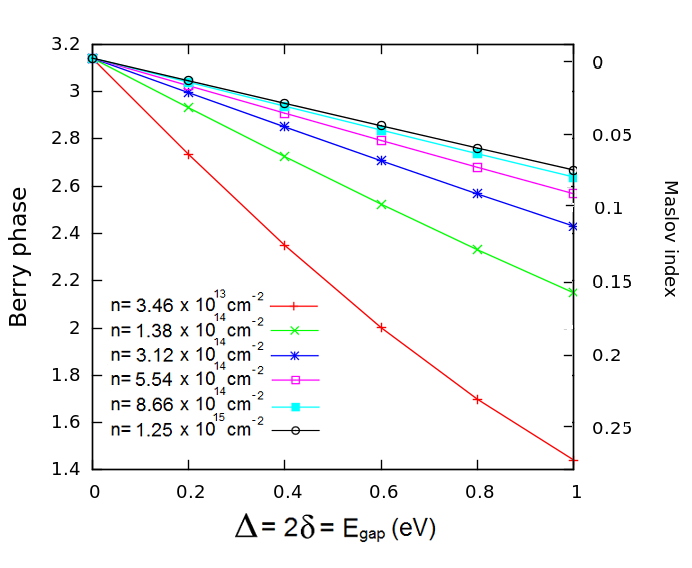}
\caption{(Color online) Berry phase and Maslov index vs. 2D charge density (top) and gap (bottom).}
\label{fig4}
\end{figure}

In Fig.\ref{fig4}, top, we report  the behavior of the phase as function of $n_{\rm 2D}$, as parametrized by $\Delta$. As $\Delta$ increases, the band structure becomes quadratic in a larger and larger $q$-space region near the former Dirac point, and the  curvature is less and less localized in $q$; in this case, a progressively larger  path radius is needed to obtain a saturated Berry phase. Clearly, however, a saturation value between 0 and $\pi$ is always reached at plausible densities, and a significant tuning of the Berry phase can obtained.  In Fig.\ref{fig4}, bottom, we report the $\Delta$ dependence of the saturation value of the phase, which appears roughly linear at least near the unperturbed limit. Also reported in these figures is the Maslov index, relevant to the magnetic spectrum, to be discussed below.

This behavior is due, in essence, to the fact that when the  perturbation removes  the conical-intersection singularity and a  quadratic dispersion appears in a finite $q$-space portion, linear sections of the dispersion do survive and give rise to a non-zero phase. This is also borne out by the circulation of the Berry connection. Let us assume that the Bloch function is a linear combination of that of  the 2DEG and of the Dirac function (Eq.\ref{diracwf})
\begin{equation}
u_{gg}={\rm e}^{i{\bf q}\cdot{\bf r}}  [a+bu_K],
\label{lincom}
\end{equation}
 with $a^2$+$b^2$=1.  For a given non-zero perturbing  potential,  both $a$ and $b$ are always smaller than 1 (except at the former Dirac point where the linear term is zero). As $q$ increases,  $a$ decreases and $b$ grows (although it never gets  back to 1 because the dispersion never recovers exact  linearity). On the other hand, at a given $q$, $a$ will dominate if the potential is large, and $b$ will in the small potential limit. The connection on the circular path or radius $q$  is $A = a^2 A_{2DEG} + b^2 A_g$, the two $A$'s being those of the 2DEG and of graphene.  The 2DEG  connection gives zero phase and the mixed terms cancel out,  so the phase is $\Gamma$=$b^2\pi$ with $b$$\in$(0,$b_{\rm max}$), where $b_{\rm max}$$<$1 depends on the potential intensity (as well as on the chosen radius $q$). If the path radius could become infinite  (i.e. the quadratic component of the dispersion disappeared entirely) 
 $b$$\rightarrow$1 and the $\pi$ phase would be recovered; but that never happens, due (at least) to the finiteness of the Brillouin zone.  The interpretations in terms of flux of the curvature and of circulation of the connection are of course consistent: as function of radius $q$,  the phase picks up strongly (Fig.\ref{fig4}) as the curvature   drops to zero (Fig.\ref{fig3}). Both quantities are signaling that electrons are leaving the quadratic-dispersion  region and  re-entering the linear-dispersion Dirac-particle-like  regime.

We now take an educated guess at the observable consequences of these results. Firstly, chirality will not be an exact quantum number for a non-zero quadratic-dispersion region \cite{castroneto}. Using the same wavefunction as for the phase (Eq.\ref{lincom}), the chirality would be $b$/2, i.e. smaller than in graphene, and growing towards (but never reaching) the graphene value for small gaps or for large path radius, i.e. large carrier density. (Of course, in this case too at some point the radius will become large enough to pick up contributions from other Dirac valleys.) It would be interesting to probe and assess chirality in a gapped system, similarly to that of graphene, even   with non-transport probes \cite{chirality-arpes}.

Secondly, the quantum Hall effect will be modified, because the Landau spectrum should now be a combination of that of free electrons and  Dirac particles. We can bypass the thicket of a direct calculation  noting that, as discussed in Refs.\cite{qheberry2,qheberry3,expmanif}, the   magnetic-quantized observables (Shubnikov-de Haas frequencies, orbit areas, Landau energies, etc.) depend on the so called Maslov index $\gamma$, which is zero for linear dispersion, and 1/2 for quadratic dispersion. The Maslov index, also displayed in Fig.\ref{fig4} above, is a property of the topology of the band structure and is related to the Berry phase $\Gamma$   \cite{qheberry3,expmanif}:
\begin{equation}
\gamma=\frac{1}{2}(1 - \frac{\Gamma}{\pi}).
\end{equation}
Since the Berry phase can be changed continuously by the gap opening as discussed here, we expect that measurable quantities in the QHE and related phenomena will take on  intermediate values between graphene and the electron gas.
 
So far we have implicitly assumed that the perturbation opening the gap has the periodicity of graphene itself (which is the case, for example, for combined shear-tension strains preserving vibrational stability \cite{gapopening3}). This consideration is relevant in view of a recent  report of quantum Hall effect measurements \cite{butterfly} for graphene on $h$-BN, which suggest that the system has a gap of order 20 meV and no $n$=0 Landau level--that is, the Berry phase would be essentially zero. This is not in contrast with our result. Because the gap-opening potential produced by epitaxy on BN has a wavelength of order 20 unit cells, the Brillouin zone shrinks to about 1/20th of its size in graphene; considering the bands in e.g. Fig.\ref{fig2}, it is clear that  in such a case the corners of the zone would be too close for a significant linear portion of the dispersion to survive (assuming, as is natural, that the Fermi velocity, i.e. ultimately the Dirac cone opening angle, does not change). So in terms of our present interpretation,  the suppression of the phase in Ref.\cite{butterfly} is not due to the gap opening itself but to the Brillouin zone collapse due to the long wavelength of the gap-opening perturbation. We  thus  expect that long-wavelength perturbing potentials will generally suppress the Berry phase more strongly, or even entirely. By the same token, the perturbing-potential wavelength offers a further adjustable parameter for the tuning of the phase. 

In summary, we have shown that, when a gap of tunable size opens at the conic band intersections of graphene, the Berry phase  does not vanish abruptly, but progressively decreases as the gap increases; the phase depends on, and increases with, the radius of the path in reciprocal space (i.e. the Fermi wave vector for a doped system). The phase and its observable consequences can thus be tuned continuously via gap opening  --by a modulating potential induced by strain, epitaxy, or nanostructuration-- and doping adjustment. The perturbation wavelength also affects the degree of suppression of  the phase,  weak at short wavelength and strong at long wavelength.

Work supported in part by MIUR-PRIN 2010 {\it Oxide}, Fondazione Banco di Sardegna, CINECA-ISCRA grants, CAR and PRID of University of Cagliari.

\end{document}